# Energy landscape and phase competition of $CsV_3Sb_5$-, $CsV_6Sb_6$-, and $TbMn_6Sn_6$-type Kagome materials

Guanghui Cai[1,2], Yutao Jiang[1,2], Hui Zhou[1,2], Ze Yu[1,2], Kun Jiang[1], Youguo Shi[1], Sheng Meng[1,2,3 *], Miao Liu[1,3,4 *]

[1] *Beijing National Laboratory for Condensed Matter Physics, Institute of Physics, Chinese Academy of Sciences, Beijing, 100190, China*

[2] *School of Physical Sciences, University of Chinese Academy of Sciences, Beijing 100190, China*

[3] *Songshan Lake Materials Laboratory, Dongguan, Guangdong, 523808, China*

[4] *Center of Materials Science and Optoelectronics Engineering, University of Chinese Academy of Sciences, Beijing, 100049, P. R. China*

*Corresponding author: smeng@iphy.ac.cn, mliu@iphy.ac.cn

Guanghui Cai and Yutao Jiang contributed equally to this work.

## Abstract

Finding viable Kagome lattices is vital for materializing novel phenomena in quantum materials. In this work, we performed element substitutions on $CsV_3Sb_5$ with space group $P6/mmm$, $TbMn_6Sn_6$ with space group $P6/mmm$, and $CsV_6Sb_6$ with space group $R\bar{3}m$, respectively, as the parent compounds. A total of 4158 materials were obtained through element substitutions, and these materials were then calculated via density function theory in high-throughput mode. Afterward, 48 materials were identified with high thermodynamic stability ($E_{hull} < 5meV/atom$). Furthermore, we compared the thermodynamic stability of three different phases with the same elemental composition and predicted some competing phases that may arise during material synthesis. Finally, by calculating the electronic structures of these materials, we attempted to identify patterns in the electronic structure variations as the elements change. This work provides guidance for discovering promising $AM_3X_5/AM_6X_6$ Kagome materials from a vast phase space.

## Introduction

The Kagome lattice[1] forms a basis for numerous unique quantum phenomena, such as spin frustration[2], unconventional superconductivity[3], Dirac/Weyl semimetals[4–6], giant anomalous Hall effect (AHE)[7–9], and charge density wave (CDW) order[10–12], due to its particular spatial configuration[13]. Thus, the discovery of synthesizable real-world Kagome compounds would significantly propel progress in the field of quantum materials. $CsV_3Sb_5$, a material containing the Kagome lattice, was previously discovered[14–16], demonstrating interesting features of flat energy bands, Van Hove singularity, and Dirac point in its electronic structure. This material was also found to exhibit superconductivity below 2.5 K[17], making it a potential quantum materials candidate, with proposed applications such as energy-efficient nano-electronic devices, as suggested by Zheng et al[18]. In addition to $\mathrm{CsV_3Sb_5}$, $\mathrm{RbV_3Sb_5}$ and $\mathrm{KV_3Sb_5}$[19][20] were identified as potential systems for exploring novel quantum phenomena, given their similar atomic and electronic structures.

Aside from compounds similar to $\mathrm{CsV_3Sb_5}$, research has revealed the synthesizable Kagome lattice-based compound $\mathrm{CsV_6Sb_6}$[21], which has a more complex bilayer structure. This finding opens up new possibilities for studying a range of topological non-trivialities in bilayer Kagome materials. Interestingly, while $\mathrm{CsV_6Sb_6}$ shares the stoichiometry of another Kagome material, $\mathrm{TbMn_6Sn_6}$[22–24], their crystal structures differ. $\mathrm{CsV_6Sb_6}$ belongs to the $\mathrm{R\overline{3}m}$ space group, while $\mathrm{TbMn_6Sn_6}$ falls into the $\mathrm{P6/mmm}$ space group. Detailed comparisons of the crystal structures of $\mathrm{CsV_3Sb_5}$, $\mathrm{CsV_6Sb_6}$, and $\mathrm{TbMn_6Sn_6}$ are depicted in Fig.1a-c.

Theoretical prediction by computer can greatly accelerate the discovery process of materials, e.g., it expedited the in silico design of the Mg ion battery[25][26] and N-containing ternary compound[27], and the theoretical predictions are all confirmed by experiments shortly. In this paper, we implement the same technique to $\mathrm{CsV_3Sb_5}$-, $\mathrm{CsV_6Sb_6}$-, and $\mathrm{TbMn_6Sn_6}$-based systems(denoted as CVS135-type, CVS166-type and TMS166-type systems respectively for convenience throughout this paper), aiming to methodically identify potential Kagome materials from these systems. This paper expands previous work[28] which only analyzed CVS135-type compounds, by investigating a broader structural phase space (three structure templates), and providing a thorough analysis of trends and insights into understanding phase competition in these compounds.

To elaborate, we initially subjected the elements in $CsV_3Sb_5$, $CsV_6Sb_6$, and $TbMn_6Sn_6$ crystal structures to substitution, resulting in 4158 possible materials. We then employed high-throughput computation to calculate the formation energy, giving us the thermodynamic stability of all these compounds. Following a comprehensive search, we identified 48 compounds with robust thermodynamic stability ($E_{hull} < 5\mathrm{meV/atom}$), which is listed in Table.1. Our findings showed that CVS135-type, CVS166-type, and TMS166-type structures have distinct preferences for certain chemical elements. For instance, several stable TMS166-type compounds emerged when the Tb site was replaced by H/Li/Na, while introducing elements like K/Cs/Ru could bring about the stability of the compound in CVS135-type and CVS166-type structures. We also find that the CVS135-type compound is more prevalent as the CVS166-type compound typically exhibits less stability. Exceptions to this include compounds such as $RbFe_6Sb_6$, $RbTc_6Sb_6$, $CsFe_6Sb_6$, $CsTc_6Sb_6$, and $CsTc_6Ge_6$. The presence of phase competitions in K-Ti-Bi, Rb-Ti-Bi, Cs-Ti-Bi, Cs-Pd-Pb compounds indicate synthesis may be more nuanced, even if these materials are theoretically stable. This paper demonstrates a standard method for identifying new synthesizable compounds swiftly and affordably, thereby shedding light on revolutionary techniques within the domain of physical science.

## Methodology

In this paper, we employed a high-throughput workflow to calculate a total of 4,158 Kagome compounds with different compositions and space groups. Specifically, we calculated 1,386 CVS135-type compounds (whose composition can be represented as $AM_3X_5$) with space group P6/mmm, 1,386 TMS166-type compounds ($AM_6X_6$) with space group P6/mmm, and 1,386 CVS166-type compounds($AM_6X_6$) with space group $R\bar{3}m$. Among them, the "A" element comes from the element set composed of purple squares in Fig.1(d), the "M" element comes from the element set composed of indigo squares, and the "X" element comes from the element set composed of orange squares and all calculations are under the same parameters as follows:

The Vienna Ab-initio Simulation Package (VASP)[29][30] code with the projector augmented wave (PAW) method[31][32] was used to realize the density functional theory (DFT) calculations[33]. We choose the Generalized Gradient Approximation (GGA) of Perdew-Burke-Ernzerh (PBE)[34] to describe the exchange and correlation energy density function. The conjugate gradient scheme was used to optimize the atomic structures. The cutoff energy was chosen to be 520 eV. The gamma-centered k-mesh was used to sample the Brillouin-zone during the calculation at the density of $100$ kpoints $Å^{-3}$. All energies convergence criterion of the self-consistency was set to $5 \times 10^{-6}$eV.

## Results and discussions

The synthesizability of a compound primarily depends on its thermodynamic stability and the kinetics during its growth process. This means that a compound thermodynamically competes with all potential phases in the same chemical space and eventually settles into the energetically favorable phases. Consequently, determining the existence probability of a compound becomes easier if the formation energies of all stable compounds are known. Thanks to computational databases like the Materials Project[35], AFLOW[36], and Atomly[37], the energy landscape of numerous chemical spaces is readily available, making it easier for us to assess the thermodynamic stability of a given compound[38–40].

In particular, the formation energy of a compound, also known as the standard enthalpy of formation, is defined as the enthalpy change when a substance forms from its pure elements under identical conditions[41]. A negative formation energy signifies energy released by the reaction and relative stability of the compound and vice versa. The most stable compounds in a chemical system delineate the lower boundary of formation energies or the energy convex hull in the chemical system, given that the formation energies of stable compounds are negative. The distance between a compound's formation energy and the energy convex hull, also referred to as "the energy above the hull ($E_{hull}$)", serves as a valuable metric for verifying the thermodynamic stability of the compound. Essentially, $E_{hull}$ denotes the driven force in energy to decompose a compound[42–44].

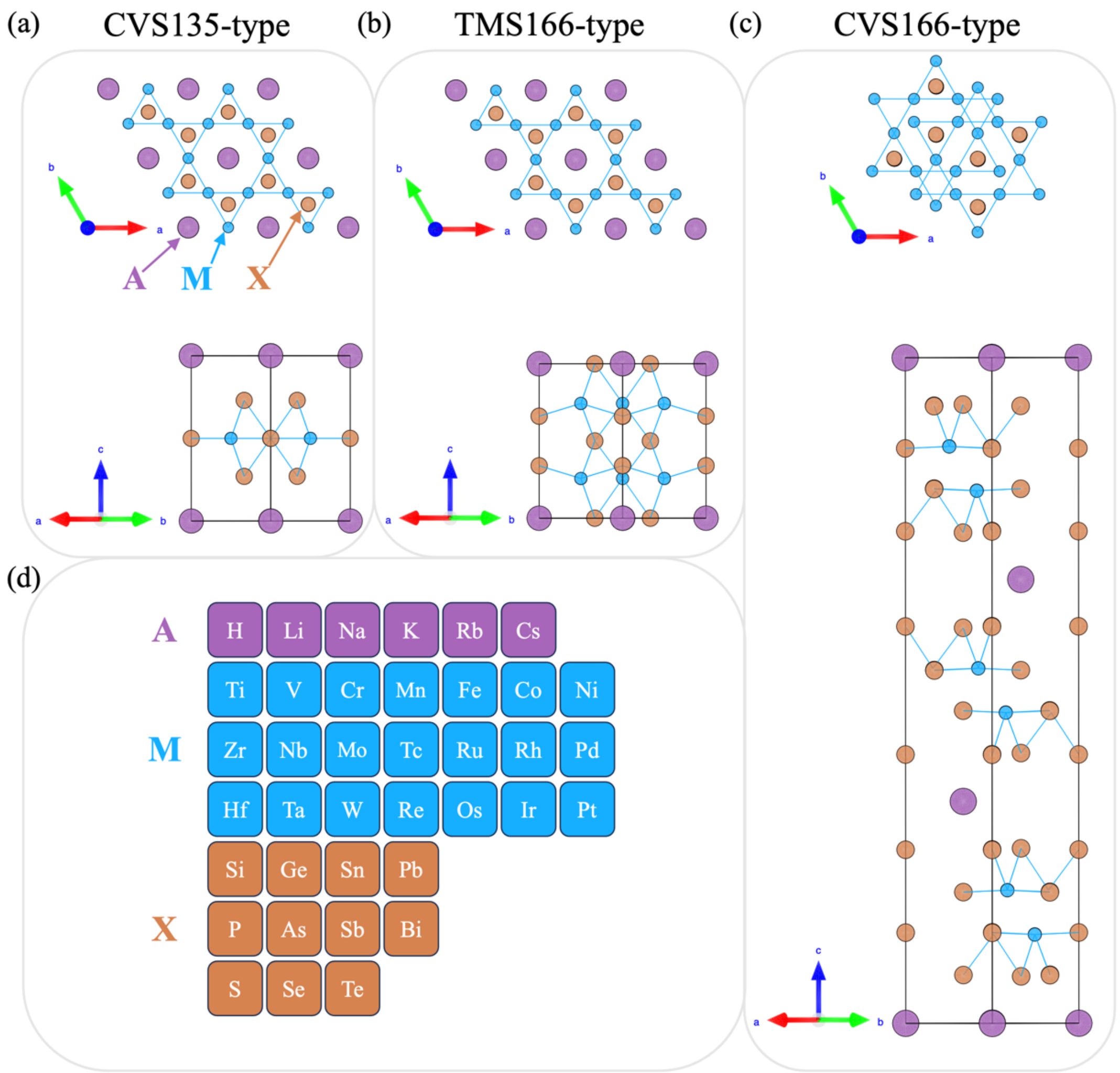

***Fig.1:*** Three types of Kagome materials have distinct structural templates: (a) $CsV_3Sb_5$-type, space group $P6/mmm$, denoted as CVS135-type structure. (b) $TbMn_6Sn_6$-type, space group $P6/mmm$, denoted as TMS166-type structure. (c) $CsV_6Sb_6$-type, space group $R\overline{3}m$, denoted as CVS166-type structure. When substituting the elements in the above three structural templates, the chosen elements are displayed in (d).

In this paper, we first expand the chemical space of the $AM_6X_6$, covering both the $R\overline{3}m$ and $P6/mmm$ structures, totaling 2772 compounds. Then the $E_{hull}$ are derived for all the compounds by plugin the data from the Atomly database. The $AM_3X_5$ (1386 structures) compounds were previously investigated by Jiang et al and these data are included for comparison[28]. All the data can be inquired in atomly.net[37].

In Fig.2, we plot the $E_{hull}$ of the three different structures mentioned in the article with the same element composition in each circle and use this to elucidate their stability and phase competition. The CVS135-type (space group $P6/mmm$) compounds are plotted in red color, whereas the CVS166-type (space group $R\overline{3}m$) compounds are shown as yellow color, and the TMS166-type (space group $P6/mmm$) compounds are plotted as blue color. The radius of each sector in the figure represents the magnitude of stability, the larger the radius stands for the higher the stability. It can be generally found from the Fig.2 that among the hydrogen-containing, lithium-containing and sodium-containing compounds, TMS166-type compounds with space group $P6/mmm$ have more thermodynamically stable structures than the other two types of Kagome structures; In the compounds containing potassium, rubidium and cesium there are almost no TMS166-type compounds, but many CVS135-type compounds with space group $P6/mmm$ and CVS166-type compounds with space group $R\overline{3}m$. Therefore, the stability of those compounds is closely element-dependent.

From the Fig.2, it can be found that the theoretical calculation is indeed in good agreement with the experiments. For example, $CsV_3Sb_5$ [16], $RbV_3Sb_5$ [45], $KV_3Sb_5$ [20] (CVS135-type compounds with space group $P6/mmm$), $LiMn_6Sn_6$ [8] (TMS166-type compounds with space group $P6/mmm$), and $KV_6Sb_6$, $RbV_6Sb_6$, $CsV_6Sb_6$ [46] (CVS166-type compounds with space group $R\overline{3}m$) are thermodynamic stable and thus synthesizable. Therefore, the mothed and treatment employed in this paper is reliable. As shown in Fig.2, there are several new compounds with good thermodynamic stability ($E_{hull} < 5meV/atom$). Those are $KTi_6Bi_6$, $RbTi_6Bi_6$, $RbFe_6Sb_6$, $RbTc_6Sb_6$, $CsTi6Bi6$, $CsFe_6Sb_6$, $CsTc_6Sb_6$, $CsTc_6Ge_6$, $CsPd_6Pb_6$ in CVS166-type structure and $HFe_6Ge_6$, $HCo_6Ge_6$, $HNi_6Sn_6$, $HRh_6Sn_6$, $LiTi_6Bi_6$, $LiFe_6Ge_6$, $LiNi_6Ge_6$, $LiNi_6Si_6$, $NaRh_6Pb_6$, $NaRh_6Sn_6$, $NaPd_6Pb_6$ in TMS166-type structure, requesting the future experimental confirmation. Table.1 lists the details of those low $E_{hull}$ Kagome materials, including the chemical formula of materials, the structure type they belong to, the ID number in Atomly database[37], the $E_{hull}$ value of materials, and whether the material has been experimentally synthesized. Among them, the materials that have not been synthesized by experiments are very promising materials that can be synthesized in the future.

The phase competition can also be spotted from Fig.2. Here we only list two out-of-phase $E_{hull}$ which are both below 5meV/atom, and the cases where the $E_{hull}$ of two different phases are both lower than 10 meV/atom can be queried in Table.S1, For example, the K-Ti-Bi in CVS166-type structure and in CVS135-type structure are both relatively stable as the $E_{hull}$ for CVS166-type structure is 3.79 meV/atom and $E_{hull}$ for CVS135-type structure is 0.0 meV/atom, only 3.79 meV/atom difference in formation energy. Hence, it adds an additional challenge for synthesizing the desired phase. Similarly, such a phase competition can also be found in compounds containing Rb-Ti-Bi, Cs-Ti-Bi and Cs-Pd-Pb elements, and they are all a competition between CVS135-type and CVS166-type structures.

Previously, we used a similar method to explore the thermodynamic stability of the CVS135-type Kagome materials[28], a couple of predicted phases such as $CsTi_3Bi_5$, $RbTi_3Bi_5$ have been successfully synthesized[47][48]. In this work, as we further expanded the chemical space of those systems, it is found that the newly added compound may have even lower formation energy and break the existing convex hull, resulting in shooting up the $E_{hull}$ to larger values. Therefore, the stable compounds found in previous work may become less stable as there is another new stable phase added. For example, $CsV_3Bi_5$ was found as the stable structure previously, but the newly added compound in Cs-V-Bi chemical phase space is found to have even smaller formation energy and is indeed the new stable compound.

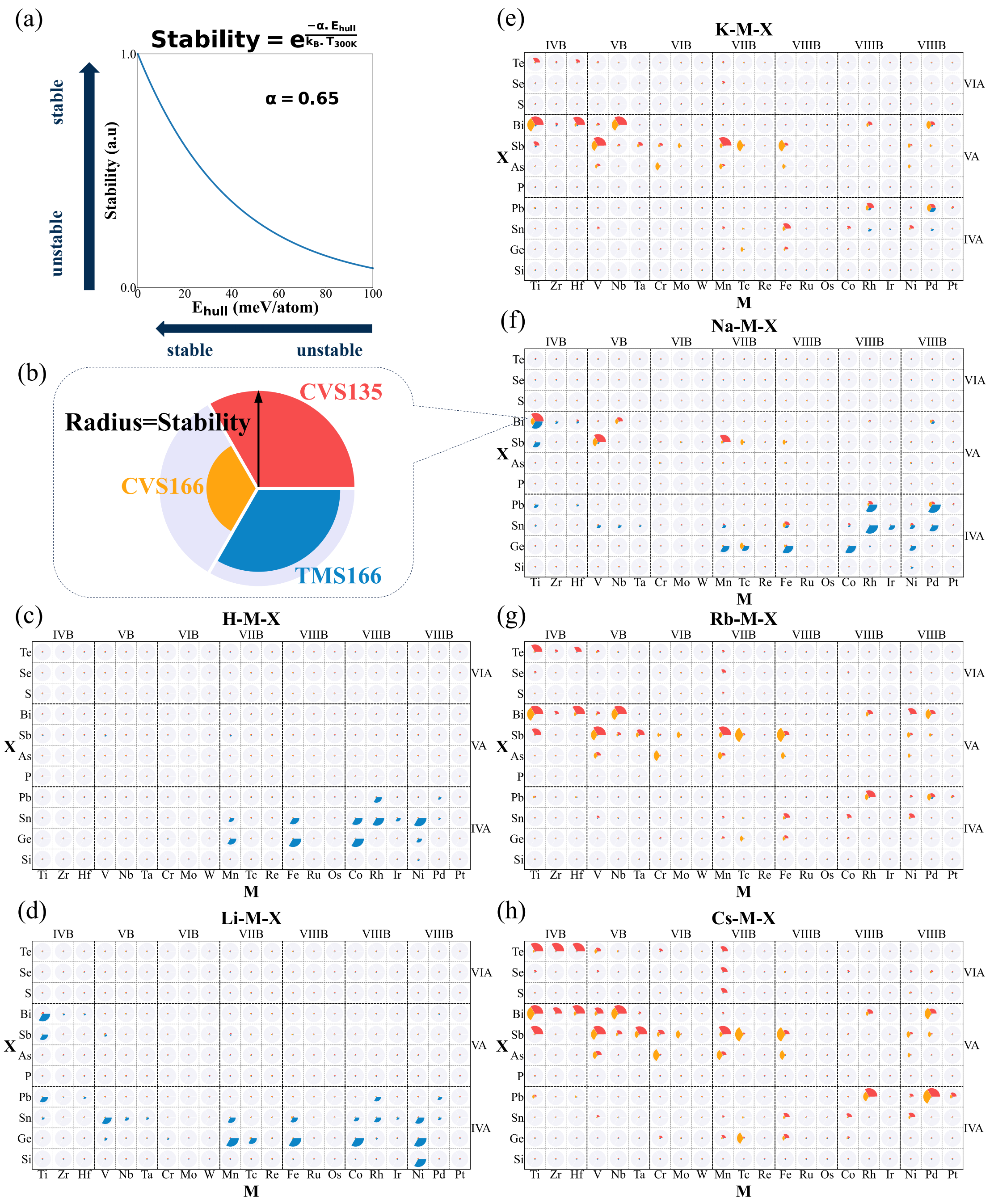

*Fig.2:* Thermodynamic stability of three types of Kagome structures in expanded chemical space. We first define the stability measure of a compound based on the $E_{hull}$ and Boltzmann distribution, as shown in (a). Stability=1.0 stands for stable compounds, stability=0.61 is corresponding to $E_{hull} = 20meV/atom$, and stability=0.08 is corresponding to $E_{hull} = 100meV/atom$. (b) The three sectors of the pie chart represent three Kagome crystalline types: the red sector represents CVS135-type compound with space group $P6/mmm$, the bule sector represents TMS166-type compound with space group $P6/mmm$, and the yellow sector represents CVS166-type compound with space group $R\overline{3}m$. The radius of each sector denotes the thermodynamic stability of each compound as defined in (a). The larger the radius of a sector, the more stable the compound is. (c)-(h) represents the thermodynamic stability of the H-M-X, Li-M-X, Na-M-X, K-M-X, Rb-M-X, Cs-M-X respectively.

*Table.1:* List of top 48 stable compounds ($E_{hull} < 5$meV/atom). The chemical formula, the structure type, id of the compound in atomly.net[37], $E_{hull}$ of the compound, and references of the experiments (if available) of those compounds are listed.

| Chemical Formula | Structure Type | Atomly ID | $E_{hull}(meV/atom)$ | Experimental Ref |
|---|---|---|---|---|
| $HFe_6Ge_6$ | TMS166 | 1000301638 | 0.00 | |
| $HCo_6Ge_6$ | TMS166 | 1000301649 | 0.00 | |
| $HNi_6Sn_6$ | TMS166 | 1000301661 | 0.68 | |
| $HRh_6Sn_6$ | TMS166 | 1000301727 | 3.35 | |
| $LiTi_6Bi_6$ | TMS166 | 1000301823 | 4.84 | |
| $LiMn_6Ge_6$ | TMS166 | 3001812095 | 0.00 | |
| $LiFe_6Ge_6$ | TMS166 | 3001812096 | 0.00 | [49] |
| $LiCo_6Ge_6$ | TMS166 | 0000060277 | 0.00 | [50] |
| $LiNi_6Ge_6$ | TMS166 | 0000043146 | 0.00 | [50] |
| $LiNi_6Si_6$ | TMS166 | 0000102969 | 2.72 | [50] |
| $NaTi_3Bi_5$ | CVS135 | 1000299281 | 0.00 | |
| $NaV_3Sb_5$ | CVS135 | 1000299291 | 2.02 | |
| $NaRh_6Pb_6$ | TMS166 | 1000302186 | 4.79 | |
| $NaRh_6Sn_6$ | TMS166 | 1000302185 | 0.00 | |
| $NaPd_6Pb_6$ | TMS166 | 1000302197 | 0.00 | |
| $KTi_3Bi_5$ | CVS135 | 1000299512 | 0.00 | |
| $KTi_6Bi_6$ | CVS166 | 1000300902 | 3.79 | |
| $KV_3Sb_5$ | CVS135 | 1000299522 | 0.00 | [15] |
| $KMn_3Sb_5$ | CVS135 | 1000299544 | 0.00 | |
| $KNb_3Bi_5$ | CVS135 | 1000299600 | 0.00 | |
| $KHf_3Bi_5$ | CVS135 | 1000299666 | 2.99 | |
| $RbTi_3Bi_5$ | CVS135 | 1000299743 | 0.00 | [48] |
| $RbTi_6Bi_6$ | CVS166 | 1000301133 | 4.08 | |
| $RbV_3Sb_5$ | CVS135 | 1000299753 | 0.00 | [15] |
| $RbMn_3Sb_5$ | CVS135 | 1000299775 | 0.00 | |
| $RbFe_6Sb_6$ | CVS166 | 1000301264 | 0.00 | |
| $RbNb_3Bi_5$ | CVS135 | 1000299831 | 0.00 | |
| $RbTc_6Sb_6$ | CVS166 | 1000301242 | 0.00 | |
| $RbHf_3Bi_5$ | CVS135 | 1000299897 | 0.00 | |
| $CsTi_3Bi_5$ | CVS135 | 1000299974 | 0.00 | [48] |
| $CsTi_6Bi_6$ | CVS166 | 1000301364 | 4.48 | |
| $CsTi_3Te_5$ | CVS135 | 1000299981 | 0.00 | |
| $CsTi_3Sb_5$ | CVS135 | 1000299973 | 0.00 | |
| $CsV_3Sb_5$ | CVS135 | 1000299984 | 0.00 | [15] |
| $CsMn_3Sb_5$ | CVS135 | 1000300006 | 0.00 | |
| $CsFe_6Sb_6$ | CVS166 | 1000301495 | 0.00 | |
| $CsFe_3Sn_5$ | CVS135 | 1000300021 | 0.00 | |
| $CsFe_3Ge_5$ | CVS135 | 1000300020 | 0.20 | |
| $CsZr_3Te_5$ | CVS135 | 1000300058 | 0.00 | |
| $CsNb_3Bi_5$ | CVS135 | 1000300062 | 0.00 | |
| $CsTc_6Sb_6$ | CVS166 | 1000301473 | 0.00 | |
| $CsTc_6Ge_6$ | CVS166 | 1000301468 | 0.00 | |
| $CsRh_3Pb_5$ | CVS135 | 1000300110 | 0.00 | |
| $CsPd_3Pb_5$ | CVS135 | 1000300121 | 0.00 | |
| $CsPd_6Pb_6$ | CVS166 | 1000301569 | 1.00 | |
| $CsHf_3Bi_5$ | CVS135 | 1000299897 | 0.00 | |
| $CsHf_3Te_5$ | CVS135 | 1000300135 | 0.00 | |
| $CsTa_3Sb_5$ | CVS135 | 1000300138 | 2.27 | |

Then we calculated the full electronic structures, including DOS and band structure, for all 4158 structures generated by replacing elements in the three structural templates mentioned above. All the data can be accessed from atomly.net[37]. This analysis not only helps in discovering novel topological materials but also guides band engineering using different metal elements as dopants[51].

To investigate the effects of different elements on different sites in $AM_6X_6$ kagome compounds with TMS166-type structures, we selected four structures with good thermodynamic stability($E_{hull} < 5meV/atom$) using a controlled variable method, namely $HCo_6Ge_6$(Atomly ID: 1000301649), $LiCo_6Ge_6$ (Atomly ID: 0000060277), $LiNi_6Ge_6$ (Atomly ID: 0000043146) and $LiNi_6Si_6$(Atomly ID: 0000102969). By comparing the electronic structures of $HCo_6Ge_6$(Fig.3a) and $LiCo_6Ge_6$(Fig.3b), we found that the A-site element has little influence on the compound's electronic structure near the Fermi level, because the A-site element's energy level is located deep within the Fermi surface and cannot have a significant impact on the bands near the Fermi level. Through a comparison of the electronic structures of $LiCo_6Ge_6$(Fig.3b) and $LiNi_6Ge_6$(Fig.3c), it was observed that $LiNi_6Ge_6$ exhibits characteristics of electron-doping in comparison to $LiCo_6Ge_6$, causing its Dirac cone and Van Hove singularity to shift closer to the Fermi surface, resulting in a corresponding downward shift in the position of the flat band. We also observed a difference in the electronic structures of $LiNi_6Ge_6$(Fig.3c) and $LiNi_6Si_6$(Fig.3d). The position of the Dirac cone at the K point in $LiNi_6Si_6$ is higher than in $LiNi_6Ge_6$. This difference can be attributed to the higher electronegativity of Si compared to Ge. The greater electronegativity of silicon causes it to attract more electrons, leading to a hole-doping effect in $LiNi_6Si_6$ and an overall upward shift of the band.

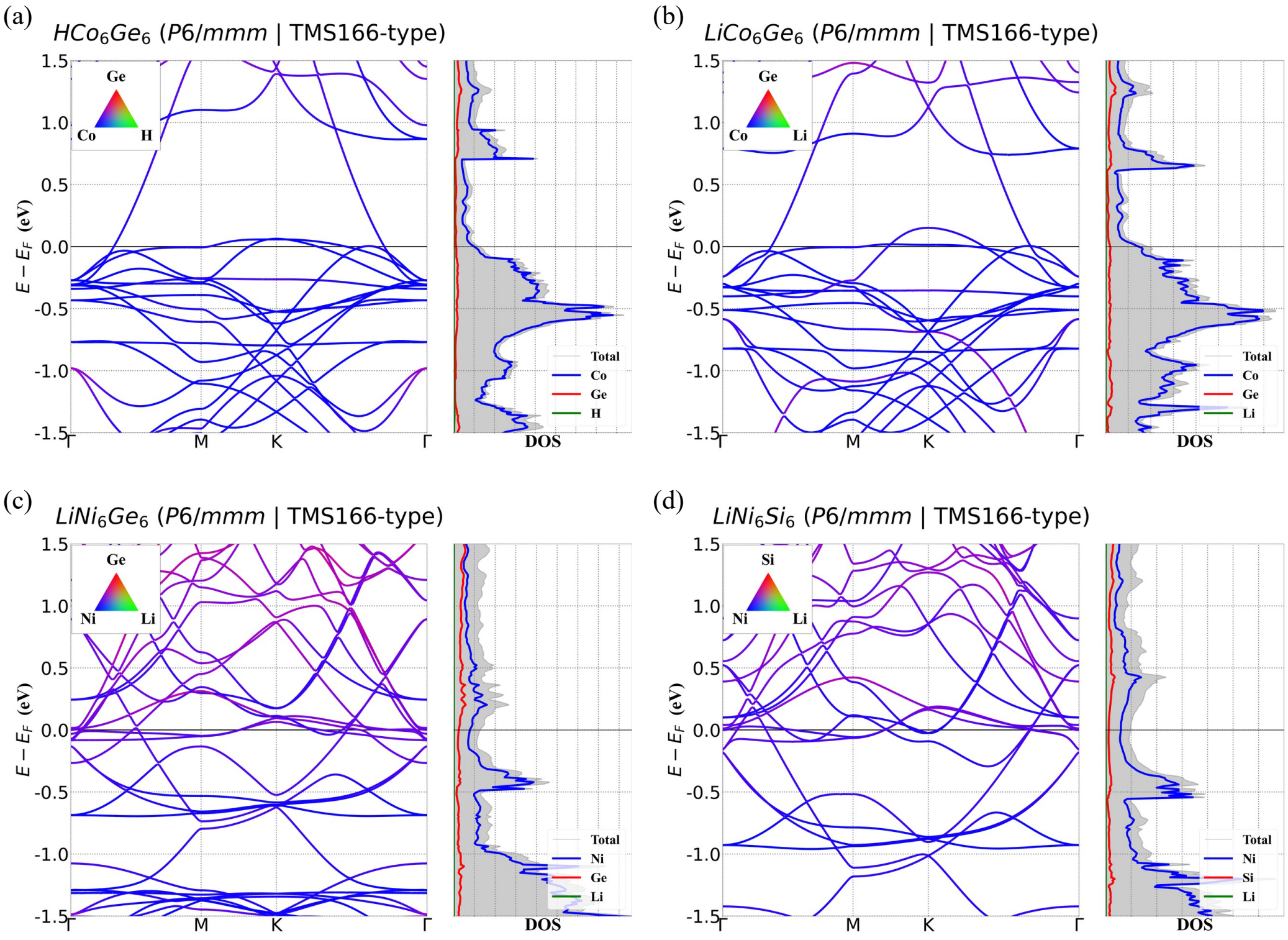

***Fig.3:*** Electronic structures of representative compounds of TMS166-type structures with $E_{hull}$<5 meV/atom. (a) the band structure of $HCo_6Ge_6$; (b) the band structure of $LiCo_6Ge_6$; (c) the band structure of $LiNi_6Ge_6$; (d) the band structure of $LiNi_6Si_6$.

The $CsV_6Sb_6$ (Atomly ID: 1000301396) compound with space group $R\overline{3}m$ exhibits a significant feature of multiple linear band crossings forming type-II Dirac nodal lines near the Fermi level in the band structure[46]. Regarding the band structures, we found that the bilayer CVS166-type compounds[52] exhibit a much simpler band topology compared to their previously studied single-layer CVS135-type counterparts[53]. Fewer bands cross the Fermi level, indicating simpler Fermi surface structures, and no ordinary Van Hove singularity[54] or type-I Dirac point like those found near the Fermi level in CVS135-type compounds. These differences may be attributed to the interlayer interactions of Kagome bilayers present in CVS166-type compounds but absent in CVS135-type compounds.

To further demonstrate the changing characteristics of the energy bands of the CVS166-type compounds($AM_6X_6$ with space group $R\overline{3}m$), we present in Fig.4 the electronic structures of representative compounds. we also observed that the A-site element has little influence on the compound's electronic structure near the Fermi level. This observation is supported by the band structure and DOS plots of all those compounds in Fig.4, which indicate that the states near the Fermi level are mostly dominated by the d orbital electrons of the M-site element, with almost no electronic state from the A-site element found near the Fermi level. So only Cs-containing compounds are shown in Fig.4. Firstly, the band structure of $CsV_6Sb_6$ (Fig.4a) exhibits multiple type-II Dirac points just above the Fermi level near F and L points, which is consistent with previous studies and verifies the accuracy of our results. Secondly, the band structures of $CsTi_6Bi_6$(Atomly ID: 1000301364) and $CsTc_6Ge_6$(Atomly ID: 1000301468) in Fig.4b and Fig.4c, respectively, can be regarded as the band structures of hole-doped and electron-doped $CsV_6Sb_6$. We note that $CsTi_6Bi_6$ exhibits more hole-like bands near the Fermi level around F and L points. However, these bands rise in energy along with the electron-like bands, resulting in less energy dispersion and the disappearance of linear band crossings near the Fermi level. For $CsTc_6Ge_6$, more electron-like bands are present near the Fermi level around F and L points, and the linear band crossings around F and L points occurs more below the Fermi level due to the higher number of valence electrons compared to $CsV_6Sb_6$, which uplifts the Fermi level. The band structures of $CsFe_6Sb_6$(Atomly ID: 1000301495) in Fig.4d shows ferromagnetism, and it exhibits linear band crossings around F and L points at approximately 0.4 eV below the Fermi level for bands of spin-down electrons exhibiting the characteristics of a Weyl point[55], but no such crossing is found in the bands of spin-up electrons, which can be used as a model systems to study interplay between topological superconductivity and magnetism[56].

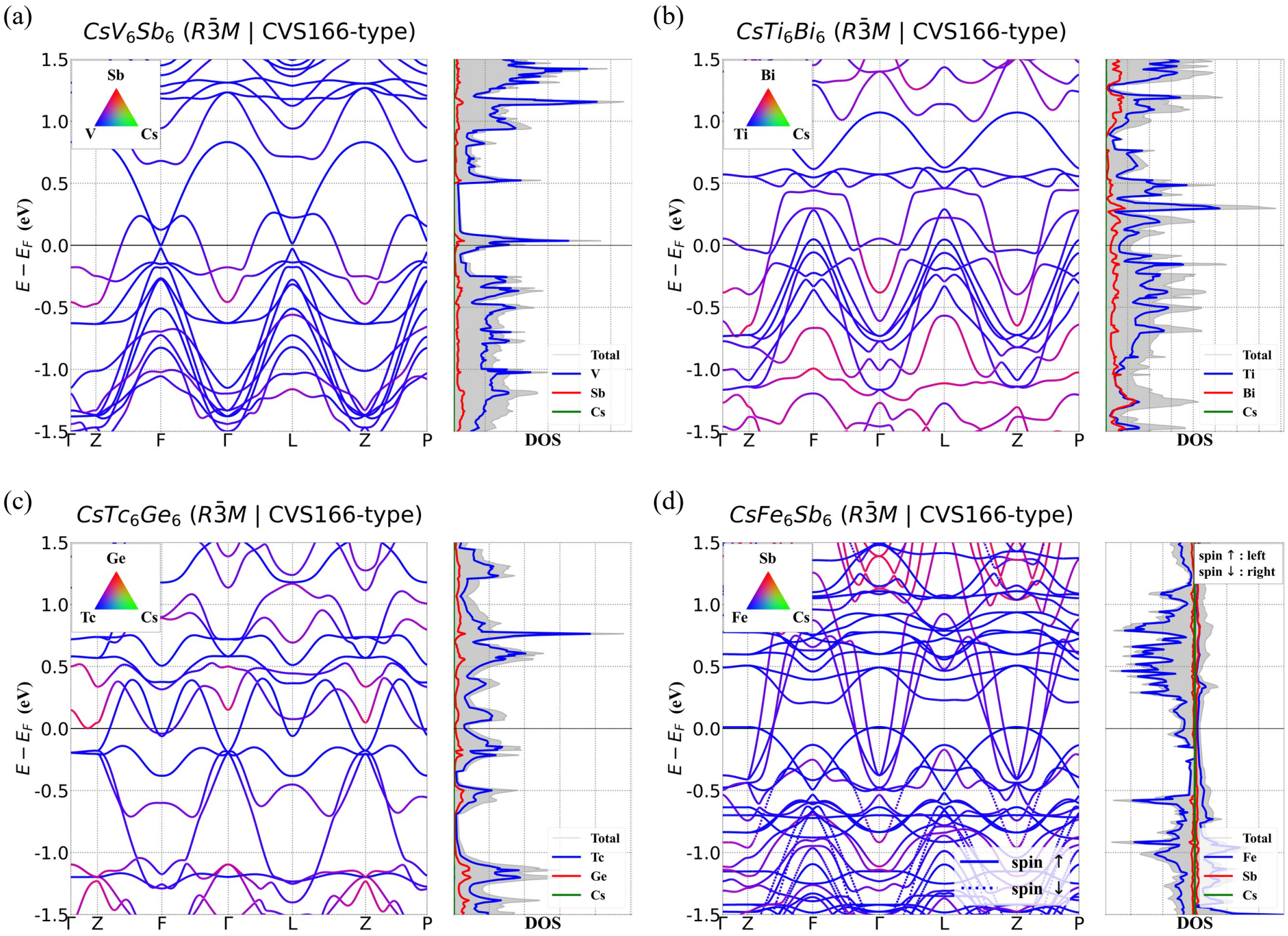

*Fig.4:* Electronic structures of representative compounds of CVS166-type structures with $E_{hull}$<5 meV/atom (The exception is the $E_{hull}$ of CaV6Sb6, which is 11.96 meV/atom, and it is shown in Table.S1). (a) the band structure of $CsV_6Sb_6$; (b) the band structure of $CsTi_6Bi_6$; (c) the band structure of $CsTc_6Ge_6$; (d) the band structure of $CsFe_6Sb_6$.

In addition to the band structure discussions, it is noteworthy that the compounds in Fig.4d with linear band crossings around F and L points, namely $CsV_6Sb_6$ and $CsFe_6Sb_6$, exhibit a nearly zero DOS at the energy where these crossings occur, a characteristic of a Weyl semimetal. Furthermore, all the band topological features of the Cs-containing compounds discussed above are also observed in their K-containing and Rb-containing counterparts. Taken together, given that $AFe_6Sb_6$ (A=K, Rb, Cs) possess similar band characteristics to $CsFe_6Sb_6$, it is highly probable that these compounds are topological semimetals with type-II nodal line fermions similar to $CsFe_6Sb_6$, with space group $R\bar{3}m$. However, to achieve precise topological classification of these compounds, additional comprehensive investigations are necessary, such as performing band symmetry group analysis, conducting band structure calculations with SOC, simulating edge states, and so on.

Consequently, it is highly anticipated that further research on the CVS166-type compounds suggested in this study will provide a more comprehensive and robust understanding of their band topology.

## Conclusions

In general, we found 13 structures, which are $HFe_6Ge_6$, $HCo_6Ge_6$, $HNi_6Sn_6$, $HRh_6Sn_6$, $LiTi_6Bi_6$, $LiMn_6Ge_6$, $LiFe_6Ge_6$, $LiCo_6Ge_6$, $LiNi_6Ge_6$, $LiNi_6Si_6$, $NaRh_6Pb_6$, $NaRh_6Sn_6$, $NaPd_6Pb_6$, with good thermodynamic stability ($E_{hull}$ less than 5meV/atom) from 1386 structures generated from the parent structure of $TbMn_6Sn_6$ with space group $P6/mmm$. Out of the 1386 structures generated from the parent structure of $CsV_6Sb_6$ with space group $R\overline{3}m$, there are 9 structures exhibiting excellent thermodynamic stability with an $E_{hull}$ less than 5 meV/atom, which are $KTi_6Bi_6$, $RbTi_6Bi_6$, $RbFe_6Sb_6$, $RbTc_6Sb_6$, $CsTi_6Bi_6$, $CsFe_6Sb_6$, $CsTc_6Sb_6$, $CsTc_6Ge_6$, $CsPd_6Pb_6$. Since some of the synthesized Kagome materials have higher $E_{hull}$, such as $LiMn_6Sn_6$(Atomly ID: 1000301858) and $CsV_6Sb_6$(Atomly ID: 1000301396), both have $E_{hull}$ greater than 5meV/atom, it is highly likely that more stable compounds can be materialized for slightly larger $E_{hull}$, which are all listed in Table.S1. We also pointed out that some materials, despite their thermodynamic stability, may pose challenges in synthesis due to the competition between different phases as these phases have almost same $E_{hull}$. It is our hope that this work can expand the scope of research on Kagome materials and lead to the discovery of more novel topological and superconducting materials.

## Acknowledgments

This research is supported by Chinese Academy of Sciences (Grant No. CAS-WX2023SF-0101, ZDBS-LY-SLH007, and XDB33020000) and the National Key R&D Program of China (2021YFA0718700). The computational resource is provided by the Platform for Data-Driven Computational Materials Discovery of the Songshan Lake laboratory.